\documentclass[conference, 10pt]{IEEEtran}
\usepackage{cite}

\ifCLASSINFOpdf
\else
\fi
\usepackage{amsmath}
\usepackage{amsmath,bm}
\usepackage{amssymb}
\usepackage{epsfig,latexsym}

\usepackage{graphicx}
\usepackage{color}

\IEEEoverridecommandlockouts

\begin{document}
%

\title{Neural Network-Based Dynamic Threshold Detection for Non-Volatile Memories}

\author{\IEEEauthorblockN{Zhen Mei, Kui Cai, and Xingwei Zhong}
\IEEEauthorblockA{Singapore University of Technology and Design (SUTD), Singapore, 487372\\
Email: $\left\lbrace \rm{mei}\_zhen, cai\_kui\right\rbrace$ @sutd.edu.sg, xingwei$\_$zhong@mymail.sutd.edu.sg}
\thanks{This work is supported by SUTD-ZJU grant ZJURP1500102, Singapore Ministry of Education Academic Research Fund Tier 2 MOE2016-T2-2-054, and SUTD SRG grant SRLS15095. 
 
This paper has been accepted by ICC 2019.}
}


%


\maketitle

\begin{abstract}
The memory physics induced unknown offset of the channel is a critical and difficult issue to be tackled for many non-volatile memories (NVMs). In this paper, we first propose novel neural network (NN) detectors by using the multilayer perceptron (MLP) network and the recurrent neural network (RNN), which can effectively tackle the unknown offset of the channel. However, compared with the conventional threshold detector, the NN detectors will incur a significant delay of the read latency and more power consumption. Therefore, we further propose a novel dynamic threshold detector (DTD), whose detection threshold can be derived based on the outputs of the proposed NN detectors. In this way, the NN-based detection only needs to be invoked when the error correction code (ECC) decoder fails, or periodically when the system is in the idle state. Thereafter, the threshold detector will still be adopted by using the adjusted detection threshold derived base on the outputs of the NN detector, until a further adjustment of the detection threshold is needed. Simulation results demonstrate that the proposed DTD based on the RNN detection can achieve the error performance of the optimum detector, without the prior knowledge of the channel.

\end{abstract}


%
\IEEEpeerreviewmaketitle

\section{Introduction}
In recent years, the solid-state non-volatile memory (NVM) technologies have been developed rapidly which offer lower power consumption, faster read access time, and better mechanical
reliability than hard disk drives (HDDs), and non-volatile data retention over DRAM and SRAM.
The current NVM market is dominated by the flash memories, while emerging NVM technologies such as the spin-torque transfer magnetic random access memory (STT-MRAM) and resistive random-access memory (RRAM) are being actively explored to be the next generation NVMs, due to their superior performance of the write/read speed, data retention time, energy consumption, endurance, and scalability \cite{yu2016emerging, chen2015portable}.

Among various noises and interferences that affect the reliability of NVMs, the memory physics induced unknown offset of the channel is a critical and difficult issue to be tackled for many NVMs. For example, in the flash memory, the charges stored in the memory cell leak away from the floating gate over time, thus causing a decrease of the memory cell threshold voltage and hence the data retention noise \cite{cai2012flash}. In the multilevel-cell phase-change memory (PCM), the structural relaxation and stress release of the phase change material cause the random fluctuation of the programmed resistances of the closely-spaced amorphous levels of the memory cell, and thereby the critical issue of ``resistance drift" for PCM \cite{papandreou2011drift}. In the recently commercialized STT-MRAM, the change of the working temperature also has a significant impact on the memory reliability. In particular, with the increase of temperature, the low resistance of the STT-MRAM cell hardly changes, but the high resistance decreases, thus leading to more overlapping of the memory resistance distributions \cite{wu2016temperature}. The corresponding deviations from the nominal values of memory readback signals ({\it e.g.} threshold voltages or resistances of memory cells), called offsets, are unknown to the channel detector, and hence will severely degrade its error performance, and lead to more decoding errors of the error correction code (ECC) subsequently.

To mitigate the unknown offset of the NVM channels, the typical techniques proposed in the literature are to estimate the NVM channel with the unknown offset periodically or when the ECC decoder fails, based on which the memory sensing thresholds ({\it i.e.} the channel detection thresholds) are adjusted accordingly \cite{lee2013estimation}. However, these techniques either require a well-predicated NVM channel model, which are difficult to be derived due to the complication of memory physics, or they assume Gaussian distribution of the memory cell readback signal, which can be non-Gaussian in practice \cite{schoeny2015analysis}. Reference cells, which are redundant cells with known stored data, are also widely applied in NVMs to estimate the unknown offset of the channel \cite{huang2014optimization}. A frequent insertion of reference cells may improve the accuracy of the detection threshold, which, however, comes at the cost of higher redundancy and thus decreasing the information storage efficiency. Moreover, similar to the data cells, the reference cells also suffer from the non-uniformity issue caused by fabrication process variations, which may lead to inaccurate estimation of the NVM channel.

Constrained coding techniques have also been proposed to improve the channel detection for NVM channels with unknown offset. Typical codes proposed in the literature include the balanced codes \cite{pelusi2015m}, and the composition check codes \cite{immink2018composition}. These codes can mitigate the unknown offset of the channel when used in conjunction with the Slepian detector. However, the corresponding code rate loss is very high. A Pearson distance detection scheme \cite{immink2014minimum}, and subsequently a dynamic threshold detection based on Pearson distance detection \cite{immink2018dynamic} are proposed recently to tackle the uncertainty of the NVM channel. These works assume that the offset for a given symbol is fixed within a codeword, which may not always hold in practice. In addition, the algorithms become unpractical for large values of the codeword length and the code alphabet.

On the other hand, in recent years the machine learning (ML) and deep learning (DL) techniques have shown amazing performance in speech recognition, natural language processing, image processing, and many other areas \cite{sutskever2014sequence}. Neural networks (NNs) have also been applied to communication systems and demonstrated superior performance from various aspects, such as the channel estimation and channel decoding \cite{ye2018power, gruber2017deep}. However, so far no much work has been reported on ML for the channel detection for NVMs. Realizing that the uncertainty of the NVM channels can be effectively tackled by using the ML and DL techniques,
in this work, we propose a novel NN-based dynamic threshold detection scheme, for NVM channels with unknown offset. We mainly use the STT-MRAM channel as an example to illustrate the proposed detection scheme, although it can also be applied to the other NVMs, such as the flash memory and PCM. The major contributions of this work are summarized as follows.
\begin{itemize}
\item[1)] We first propose novel NN detectors, which can effectively tackle the unknown offset of the NVM channel. We find that the recurrent neural network (RNN) detector outperforms the multilayer perceptron (MLP) detector, and approaches the performance of the optimum detector with the full knowledge of the channel. It also requires much smaller size of training data, and can learn the NVM channel uncertainty much faster than the MLP detector.

\item[2)] To avoid the significant increase of the read latency and power consumption incurred by the NN detectors, we further propose a novel dynamic threshold detector (DTD), whose detection threshold can be derived based on the outputs of the proposed NN detectors. Simulation results demonstrate that the DTD based on the RNN detection can achieve the error performance of the optimum detector, without the prior knowledge of the channel.

\item[3)] We propose to only activate the NN-based detection when the ECC decoder fails, or periodically when the system is in the idle state. Thereafter, the threshold detector will still be adopted by using the adjusted detection threshold derived based on the outputs of the NN detector, until a further adjustment of the detection threshold is needed. Thus leading to a significant reduction of the read latency and power consumption.

\end{itemize}


\section{Channel Model}
\begin{figure}[t]
\centering
\includegraphics[height=2.1in,width=2.74in]{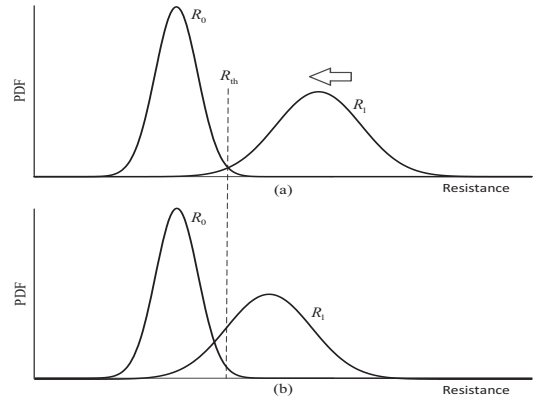}
\caption{Resistance distributions of the STT-MRAM cell. (a) Original resistance distributions; (b) With offset caused by increase of temperature.}
\label{channel}
\end{figure}

We use the STT-MRAM channel as an example to illustrate the proposed detection schemes. An STT-MRAM cell has two resistance states, a low resistance state $R_0$ which represents an input information bit of ``0'', and a high resistance state $R_1$ which denotes an information bit of ``1''. The reliability of the data stored in the memory cell is largely affected by the process variation caused by the fabrication imperfection, which leads to widened distributions of the low and high resistances of the memory cell and their overlapping, and hence the channel detection errors \cite{kui2017cascaded, mei2018magn}.
Moreover, the resistance distributions of the STT-MRAM cell are also affected by the working temperature. It has been found that with the increase of temperature, the high resistance $R_1$ will decrease and while the low resistance $R_0$ hardly changes \cite{wu2016temperature}. Fig. \ref{channel} shows an illustration of the resistance distributions ({\it i.e.} the probability density functions (PDFs) of the resistances) of STT-MRAM and their variation caused by the change of temperature. Obviously more memory sensing errors or detection errors will occur if the detection threshold $R_{\text{th}}$ remains the same when temperature increases.
Based on the stochastic characteristics of $R_0$ and $R_1$ described above, the resistance read back from the $k$-th memory cell can be expressed as
\begin{equation} \label{model}
y_{k}=r_{k}+n_{k}+b_{k},
\end{equation}
where $r_{k}$ is the nominal resistance value corresponding to an input bit of $x_k \in \{0,1\}$ stored in the $k$-th memory cell, with $k=1,\cdots,N$. That is, $r_{k}=\mu_0$ for $x_k=0$, and $r_{k}=\mu_1$ for $x_k=1$.  Here, we use $n_{k}$ to represent the variation of the resistances $R_0$ and $R_1$ caused by process imperfection, where $n_{k} \in \mathbb{R}$ is a zero-mean independent and identically distributed (i.i.d) noise sample with a variance of $\sigma^2_i$, $i=0,1$. Note that $n_{k}$ is not necessarily to be Gaussian distributed. Furthermore, we use $b_k$ to denote the offset of resistance caused by the increase of temperature, which only occurs with the high resistance state $R_1$. Thus, $b_k=0$ for all $x_k=0$. Since the influence of temperature on each cell is random, we assume the offset $b_k$ for $x_k=1$ follows a Gaussian distribution $\mathcal{N}(\mu_{b}, \sigma^{2}_{b})$ with mean of $\mu_{b}$ and standard deviation of $\sigma_b$.

In the simulations of this work, we follow the empirical results of \cite{zhang2011stt} and assume $\mu_0=1\ k\Omega$, $\mu_1=2\ k\Omega$. We further assume $\sigma_0/\mu_0=\sigma_1/\mu_1$ due to the characteristics of memory fabrication process. We vary $\sigma_0/\mu_0$ (and hence $\sigma_1/\mu_1$) and the offset $b_k$ for $x_k=1$ to account for the influence of different levels of process variations as well as the temperature increase.

\section{Neural Network-Based Dynamic Threshold Detection}

\subsection{Neural Network Detector}
In this work, we first consider the detection of the NVM channel with unknown offset as a ML problem and propose novel NN detectors. The inputs to the NN are the resistance values $\bm{y}=\left\lbrace y_1, y_2, \cdots, y_N \right\rbrace $ read back from the memory cells, where $N$ is the number of neurons in the input layer of the NN. Note that in the practical NVMs, these resistances need to be quantized first before sending to the NN. In this work, we find that by using a three or four bits uniform quantizer, the proposed detectors can achieve a performance close to that using the full soft resistances. The outputs of the NN are the soft estimates $\bm{\tilde{x}}$ of $\bm{x}$, with $\bm{\tilde{x}}=\left\lbrace \tilde{x}_1, \tilde{x}_2, \cdots, \tilde{x}_{N} \right\rbrace $, based on which we can obtain the hard estimation $\bm{\hat{x}}$ of $\bm{x}$. The corresponding hard-decision rule is: if $\tilde{x}_{k}>0.5$, $\hat{x}_k=1$; Otherwise, $\hat{x}_k=0$. We note that the NN output is a function of the NN input and the network parameters $\bm{\theta}$, given by $\bm{\tilde{x}}=f(\bm{y}, \bm{\theta})$. The NN will learn to find the best $\bm{\theta^*}$ by minimizing a properly defined loss function $\mathcal{L}$ over the set of training data, such that
\begin{equation}
\bm{\theta^*}=\text{arg}\min_{\bm{\theta}}\mathcal{L}(\bm{x},\bm{\tilde{x}}),
\end{equation}
where $\mathcal{L}(\bm{x},\bm{\tilde{x}})$ calculates the loss between $\tilde{\bm{x}}$ and $\bm{x}$. Note that the training process is to be carried out off-line. Moreover, another separate data set named the validation set will be used to validate the effectiveness of the trained NN detector. After the training and validation processes, the NN with the trained $\bm{\theta^*}$ will be applied to detect the unknown channel outputs by using the same NN architecture.

\subsubsection{Neural Network Architectures}
\begin{figure}[t]
\centering
\includegraphics[height=2.3in,width=2.8in]{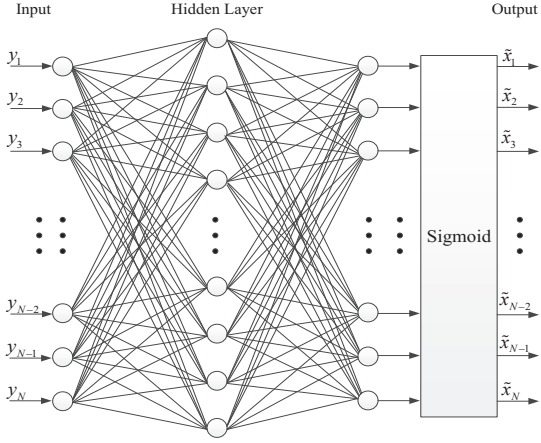}
\caption{Proposed MLP network architecture for the NN-based detection.}
\label{MLP}
\end{figure}

\begin{figure}[t]
\centering
\includegraphics[height=2.2in,width=3.1in]{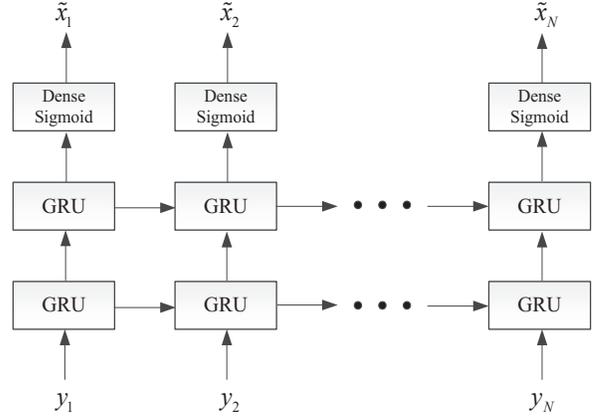}
\caption{Proposed RNN architecture for the NN-based detection.}
\label{RNN}
\end{figure}
We adopt two typical NN architectures, the MLP and RNN, to perform the NN-based detection. The MLP is a feedforward NN with fully-connected layers \cite{goodfellow2016deep}. For each neuron of the MLP, all of its weighted inputs and the bias are added up, after which an activation function $\sigma(\cdot)$ is applied to introduce the non-linearity to the NN. In this work, we adopt the rectified linear unit (ReLU) and the sigmoid activation function, which are defined by
$\sigma_{\text{relu}}(t)=\max\left\lbrace 0,t \right\rbrace$ and  $\sigma_{\text{sigmoid}}(t)=\frac{1}{1+e^{-t}},$ respectively, with $\sigma_{\text{relu}}(t)\in [0, \infty)$ and $\sigma_{\text{sigmoid}}(t)\in (0, 1)$. The proposed MLP structure is illustrated by Fig. \ref{MLP}. It consists of three layers: an input layer of size $N$, a hidden layer of size $4N$, and an output layer of size $N$. With the sigmoid function, the output $\tilde{x}_k$ of the final layer is a value between 0 and 1, which indicates the probability of $x_k$ being a `0' or a `1'.

\par Unlike the feedforward NNs, the RNN has memories to process a sequence of inputs, and hence has shown superior performance for the time series tasks. The RNN has different types of cells such as the vanilla RNN, long short-term memory (LSTM), and gated recurrent unit (GRU). Compared with the LSTM and GRU, the vanilla RNN has significantly less number of parameters. However, it suffers from the vanishing gradient problem, which causes difficulties to learn long-distance relationships since the gradients might vanish to zero \cite{goodfellow2016deep}. Therefore, in this work, we employ the GRU as the RNN unit since it can avoid the vanishing gradient problem, and it has less number of parameters than the LSTM. In our settings, we use a stacked RNN architecture as shown by Fig. \ref{RNN}. The proposed RNN has two GRU hidden layers with a many-to-many (multiple inputs, multiple outputs) configuration. The final output layer is a fully-connected layer with the sigmoid activation function. Although different codeword lengths of ECCs have been investigated for STT-MRAM \cite{mei2018magn, zhong2018rate}, in our experiments, $N$ is set to be 71, which is the same with the codeword length of the (71, 64) Hamming code adopted by Everspin’'s 16Mb MRAM \cite{mram}.

\subsubsection{Training Method}
To train the MLP and RNN, with the channel model given by \eqref{model}, we can generate sufficient number of samples of the memory readback resistance $y_k$ and its corresponding label $x_{k}$ as the training data set. In our paper, such training data set is generated for each resistance variation and offset level. We further define the specific loss function for the NNs. When the loss function is minimized through the training process, the NN output $\tilde{\bm{x}}$ will be closest to the expected output $\bm{x}$. For both the MLP and RNN, we use the mean square error (MSE) to measure the loss. Hence, the loss function is given by
$\mathcal{L}(\bm{x},\bm{\tilde{x}})=\frac{1}{N}\sum_{k=1}^{N} (x_k-\tilde{x}_k)^2$. By using variants of the gradient descent algorithm as well as the back propagation method, the optimal $\bm{\theta^*}$ can be obtained by minimizing $\mathcal{L}(\bm{x},\bm{\tilde{x}})$ defined above over the training data set, respectively.

The MLP and RNN settings obtained based on our experiments are illustrated by Table \ref{table1}. Observe that the number of network parameters for the proposed MLP and RNN is similar. The significant difference between the MLP and RNN settings is the number of training samples. After many trials, we find that $4\times 10^{4}N$ training samples are sufficient for the RNN to achieve its best performance, while the size of the training data required by the MLP is 25 times larger than the RNN. This is a great advantage of the RNN, since the size of the training data is often limited in the practical applications.

\renewcommand\arraystretch{1.5}
\begin{table}[t]
\centering
\caption{Network Settings for the Proposed MLP and RNN Architectures for the NN-based Detection.}
\begin{tabular}{|c|c|c|}
\hline
                 & MLP                                               & RNN
 \\ \hline
Network Parameters & $40683$ & $46080$
 \\ \hline
Training Samples & $1\times 10^{6}N$ & $4\times 10^{4}N$ \\ \hline
Mini-batch Size  & $4N$                                                & $2N$                                                 \\ \hline
Loss Function    & MSE                                               & MSE                                               \\ \hline
Initializer      & Xavier uniform                        & Xavier uniform                         \\ \hline
Optimizer        & Adam                                              & Adam                                              \\ \hline
\end{tabular}
\label{table1}
\end{table}

\par
\begin{figure}[t]
\centering
\includegraphics[height=0.43\columnwidth]{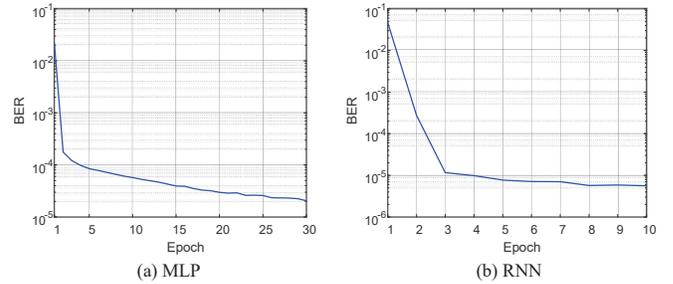}
\caption{BERs of the MLP detector and the RNN detector for each epoch during training.}
\label{train_ber}
\end{figure}

 To illustrate the NN training process, we show in Fig. \ref{train_ber} the bit error rate (BER) of the MLP detector and the RNN detector for each epoch during training. The corresponding channel parameters are： $\sigma_0/\mu_0=5\%$, $\mu_b=-0.2\ k\Omega$, and $\sigma_{b}/\mu_1=4\%$. We observe that the training BER of both NN detectors decreases as the epoch increases, and after a certain number of epoches, the BER converges. Furthermore, the MLP detector requires much more number of epoches, and the BER converges much slower than the RNN detector. This indicates that the RNN can learn the NVM channel uncertainties much faster than the MLP. Note that with preprocessing of the input data and regularization techniques, it is possible to further improve the performance of MLP. For a fair comparison, these techniques are not included in this paper.

We finally remark that the above proposed NNs and the learning process can be efficiently implemented in parallel with low-precision data types on a graphical processing unit (GPU) or an application specific integrated circuit (ASIC). The NNs after training and validation can then be used to detect its input samples $\bm{y}$ and generate the estimation of the channel input $\bm{\tilde{x}}$ by using the same NN architecture, without any prior knowledge of the NVM channel.

\subsection{Dynamic Threshold Detector Based on Neural Network Detection}
As what will be shown in Section IV, for the NVM channel with unknown offset, the above proposed NN detectors and especially the RNN detector can achieve performance very close to the optimum detector with the full knowledge of the channel. However, the corresponding NN-based detection needs to be activated for each data block of length $N$. This will lead to a significant delay of the read latency and more power consumption. Therefore, in this subsection, we propose a novel dynamic threshold detector (DTD) whose detection threshold is derived based on the outputs of the proposed NNs.

First, for a given $\bm{y}$ and with an assumed detection threshold $R_\text{th}$, we can obtain the hard estimation $\bm{\bar{x}}_{R_\text{th}}$. Therefore, based on the output $\bm{\tilde{x}}$ and hence $\bm{\hat{x}}$ from the proposed NNs, an adjusted detection threshold $R^{\text{adj}}_{\text{th}}$ can be obtained by searching for an $R_\text{th}$ that  minimizes the Hamming distance between $\bm{\hat{x}}$ and $\bm{\bar{x}}_{R_\text{th}}$, denoted by $d(\bm{\hat{x}}, \bm{\bar{x}}_{R_\text{th}})$. By including a large amount $M$ of NN output sequences, a more accurate adjusted detection threshold can be obtained. We thus have
\begin{equation} \label{R_pred}
R^{\text{adj}}_{\text{th}}=\text{arg}\min_{R_\text{th}}\sum_{i=1}^{M} d(\bm{\hat{x}}^{i}, \bm{\bar{x}}^{i}_{R_\text{th}}).
\end{equation}

Note that the above described NN-based detection and the subsequent search of the adjusted detection threshold only need to be invoked when the ECC decoder fails, or periodically when the system is in the idle state, and will be terminated once the adjusted detection threshold is determined. Thereafter, the conventional threshold detector will be adopted by using the adjusted detection threshold until a further adjustment of the detection threshold is needed. Thus leading to a significant reduction of the read latency and power consumption compared to the NN detectors described in the previous subsection, which need to be activated for every input data block.

\subsection{Optimum Detection with Full Knowledge of the Channel}

Finally, in order to provide references for evaluating the performance of the above proposed detectors, in this subsection, we derive the optimal detection threshold $R_{\text{th}}^{\text{opt}}$ and the BER of the corresponding threshold detector. We consider three cases: channels with no offset, channels with a fixed offset of $b_k=\mu_b$ ($\sigma_b=0$) for all $x_k=1$ corresponding to the high resistance state $R_1$, and channels with an offset $b_k$ that varies from cell to cell for $x_k=1$. We assume that the channel given by (1) including the knowledge of $b_k$ is known to the detector. Based on the hard-decision rule, and for a given $R_{\text{th}}$ and $b_{k}$, the BER of the threshold detector is given by
\begin{align}\nonumber \label{ins_ber}
  P_b(R_{\text{th}}, b_{k})&=\text{Pr}(x_{k}=0)\text{Pr}(\hat{x}_{k}\neq 0|x_{k}=0)\\  \nonumber
  &+\text{Pr}(x_{k}=1)\text{Pr}(\hat{x}_{k}\neq 1|x_{k}=1) \\
    & = \frac{1}{2}\left( 1+Q\left(\frac{R_{\text{th}}-\mu_0}{\sigma_0}\right) -Q\left(\frac{R_{\text{th}}-\mu_1-b_{k}}{\sigma_1}\right)   \right),
\end{align}
where we assume i.i.d channel inputs, and the resistance variation $n_k$ is Gaussian distributed. The corresponding optimal $R_{\text{th}}$ can be obtained by minimizing \eqref{ins_ber}. That is, the derivative of $P_b(R_{\text{th}}, b_{k})$ with respect to $R_{\text{th}}$ is given by
\begin{align} \nonumber
P'_b(R_{\text{th}}, b_{k})&=-\frac{1}{2\sigma_0\sqrt{2\pi}}\exp\left( -\frac{(R_{\text{th}}-\mu_0)^2}{2\sigma^{2}_0} \right) \\
& +  \frac{1}{2\sigma_1\sqrt{2\pi}}\exp\left( -\frac{(R_{\text{th}}-\mu_1-b_{k})^2}{2\sigma^{2}_1} \right) .
\end{align}
Hence, the optimal $R_{\text{th}}$ that minimizes $P_b(R_{\text{th}}, b_{k})$ can be derived by solving  $P'_b(R_{\text{th}}, b_{k})=0$ analytically, and the obtained optimum threshold is given by \eqref{rth}.

\begin{figure}[t]
\centering
\includegraphics[height=2.2in,width=3.5in]{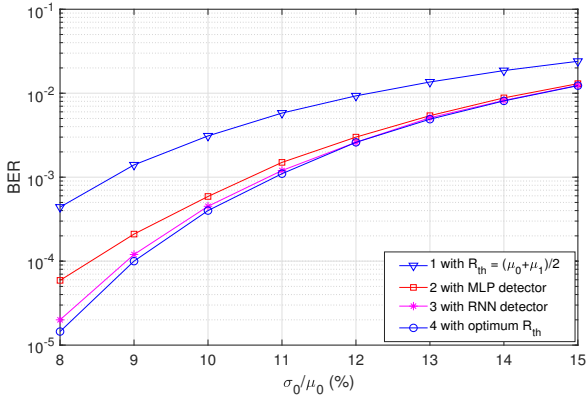}
\caption{BER comparison of different detectors for the channel without offset.}
\label{non_offset}
\end{figure}

\begin{figure}[t]
\centering
\includegraphics[height=2.2in,width=3.5in]{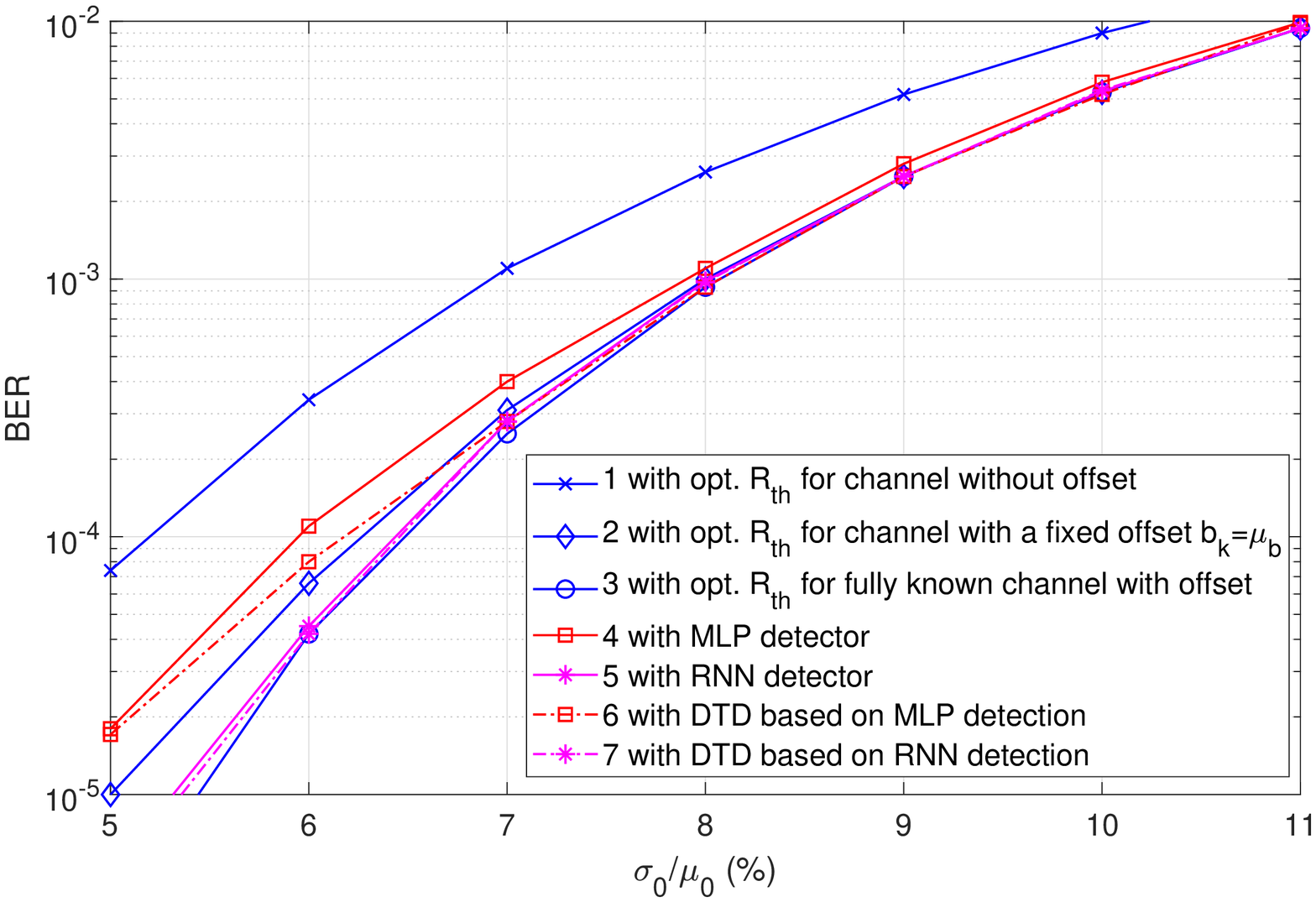}
\caption{BER comparison of different detectors for the channel with an offset of $\mu_b=-0.2\ k\Omega$ and $\sigma_{b}/\mu_1=4\%$.}
\label{offset_04}
\end{figure}

\begin{figure*}
\begin{equation} \label{rth}
R_{\text{th}}^{\text{opt}}=\frac{(\mu_1+b_k)\sigma_{0}^{2}-\mu_0\sigma_{1}^{2}-\sigma_{0}\sigma_{1}\sqrt{(\mu_0-\mu_1-b_k)^2+2\ln\frac{\sigma_{0}}{\sigma_{1}}(\sigma_{0}^{2}-\sigma_{1}^{2})}}{\sigma_{0}^{2}-\sigma_{1}^{2}},
\end{equation}
\hrule
\end{figure*}

\begin{figure}[t]
\centering
\includegraphics[height=2.2in,width=3.5in]{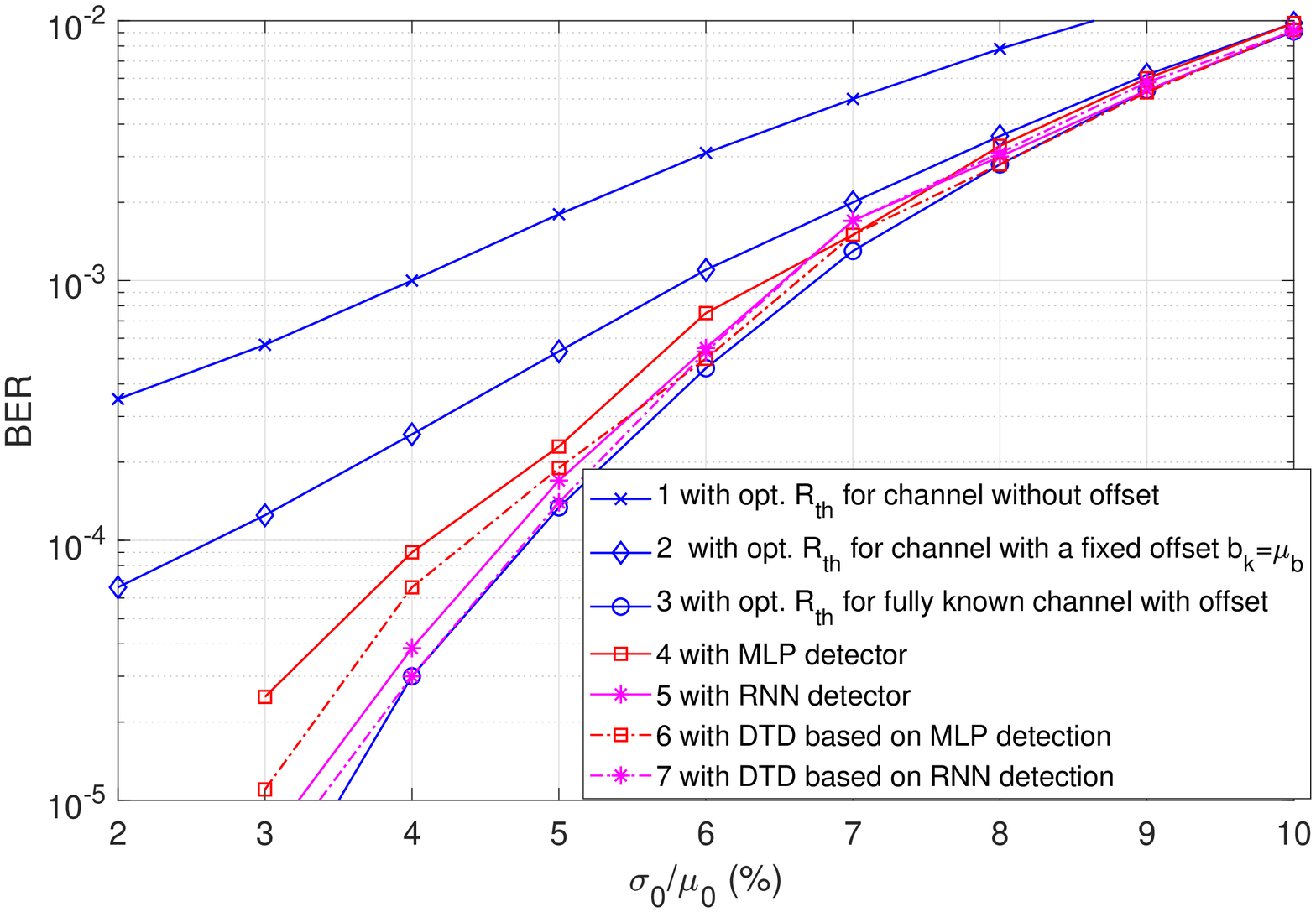}
\caption{BER comparison of different detectors for the channel with an offset of $\mu_b=-0.2\ k\Omega$ and $\sigma_{b}/\mu_1=7\%$.}
\label{offset_07}
\end{figure}

\begin{figure}[t]
\centering
\includegraphics[height=2.2in,width=3.5in]{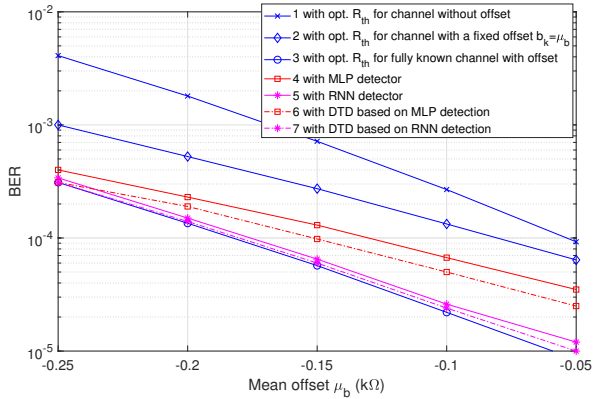}
\caption{BER comparison of different detectors for the channel with different mean offsets $\mu_b$, with $\sigma_{b}/\mu_1=7\%$ and $\sigma_0/\mu_0=5\%$.}
\label{dif_offset}
\end{figure}

\begin{figure}[t]
\centering
\includegraphics[height=2.2in,width=3.5in]{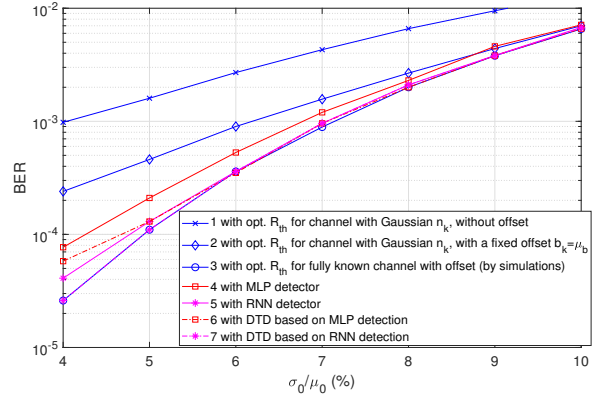}
\caption{BER comparison of different detectors for the channel with $n_k$ being Beta distributed, $\mu_b=-0.2\ k\Omega$, and $\sigma_{b}/\mu_1=7\%$.}
\label{beta}
\end{figure}

For channels with no offset, or with a fixed offset of $b_k=\mu_b$, the minimum BER of the channel detector can be obtained by substituting the optimized $R_{\text{th}}^{\text{opt}}$ of \eqref{rth} back into \eqref{ins_ber}. For the case that the offset $b_k$ has variations, the BER for a given $R_{\text{th}}$ can be obtained
by calculating the expectation of $P_b(R_{\text{th}}, b_{k})$ given by \eqref{ins_ber}. We thus have
\begin{align} \label{pb_var}
P_b(R_{\text{th}}) &= \frac{1}{2}\left( 1+ Q\left(\frac{R_{\text{th}}-\mu_0}{\sigma_0}\right) - \mathbb{E}\left[ Q\left(\frac{R_{\text{th}}-\mu_1-b_{k}}{\sigma_1}\right) \right] \right)    ,
\end{align}
where the expectation term in \eqref{pb_var} can be computed as $\mathbb{E}\left[ Q\left(\frac{R_{\text{th}}-\mu_1-b_{k}}{\sigma_1}\right) \right]=\int_{-\infty}^{\infty}p(b_{k})Q\left(\frac{R_{\text{th}}-\mu_1-b_{k}}{\sigma_1}\right)db_{k}$, with $p(b_{k})$ being the PDF of $b_k$. Since there is no close-form solution for the derivative of $P_b(R_{\text{th}})$ given by \eqref{pb_var}, we calculate it  numerically. We apply, for example, the bisection
searching method to find the root $R_{\text{th}}^{\text{opt}}$ that minimizes \eqref{pb_var}, and hence can obtain the minimum BER thereafter. The above derived minimum BERs for various cases serve as lower bounds to evaluate the performance of the proposed detectors.

\section{Performance Evaluations}

In our experiments, the implementation and training of all NNs are performed by using the machine learning library Keras \cite{keras}, with TensorFlow \cite{tensor} as its back-end. The network settings are given in Table \ref{table1}. To evaluate the BER performance of the MLP detector and RNN detector as low as $10^{-5}$, we set the test data size of $10^{6}N$ bits. In the simulations, we adopt the channel model of (1), and take different values of $\sigma_0/\mu_0$ (and hence $\sigma_1/\mu_1$) and the offset $b_k$ for $x_k=1$ to incorporate the influence of different process variations as well as the change of working temperature. The resistance variation $n_k$ is assumed to be Gaussian distributed for most cases, and the corresponding performance bounds with the optimum detection thresholds we derived in Section III.C are included as references. Meanwhile, we also present at the end of this section a case that $n_k$ is not Gaussian distributed.

To verify the effectiveness of the proposed NN detectors, we first consider the case that the  channel has no offset. As shown by Fig. \ref{non_offset},  both our proposed NN detectors significantly outperform the conventional threshold detector with the detection threshold of $(\mu_0+\mu_1)/2$, and achieve performance close to the optimum detection for the channel without offset (Curve 4). In particular, the performance of the RNN detector approaches that of the optimum detector, while the MLP detector has a larger gap from the optimum detector.

 Next, we present the performance of the various detectors for the channel with different offsets. We first illustrate in Figures \ref{offset_04} and \ref{offset_07}, the detector performance for the offsets with a fixed mean value of $\mu_b=-0.2\ k\Omega$, and different normalized root mean squared values $\sigma_{b}/\mu_1$ of $4\%$ and $7\%$, respectively. For the case of the offset with a small variation of $\sigma_{b}/\mu_1=4\%$, we observed from Fig. 6 that both the MLP and RNN detectors outperform the detector optimized for the channel without offset (Curve 1). The performance of MLP detector is slightly worse than the detector optimized for the channel with a fixed offset of $b_k=\mu_b$ (Curve 2, assuming only the mean of $b_k$ is known to the detector). However, the RNN detector provides performance better than Curve 2 and approaches that of the detector optimized using the full knowledge of the channel ({\it e.g.} $\mu_b=-0.2\ k\Omega$ and $\sigma_{b}/\mu_1=4\%$) as indicated by Curve 3. Furthermore, the proposed DTDs based on the MLP detector and the RNN detector both achieve even better performance than their original NN detectors. The DTD based on the RNN detection almost achieves the performance of the optimum detector with the full knowledge of the channel.

 With the increase of the offset variation, as shown by Fig. \ref{offset_07}, the performance gap between the detector optimized for the channel with a fixed offset of $b_k=\mu_b$ (Curve 2) and that optimized for the fully known channel (Curve 3) becomes larger. In this case, all the proposed detectors (Curves 4, 5, 6, 7) achieve performance better than the detector optimized for the channel with a fixed offset of $b_k=\mu_b$. Again, the RNN detector outperforms the MLP detector and achieves BERs close to those of the optimum detector. Moreover, the proposed DTDs outperform their original NN detectors, and the RNN detection based DTD achieves the performance of the optimum detector with the full knowledge of the channel.

 We further illustrate by Fig. \ref{dif_offset}, the performance of various detectors with different mean offsets $\mu_b$. The resistance spread is fixed at $\sigma_0/\mu_0=5\%$, and the normalized offset variation is $\sigma_{b}/\mu_1=7\%$. Observe that all the proposed detectors outperform the detector optimized for the channel with a fixed offset of $b_k=\mu_b$. The RNN detector performs better than the MLP detector, and the DTD based on the RNN detection almost achieves the performance of the optimum detector, for all the different mean offsets.

Finally, we consider the case that the resistance variation $n_k$ is non-Gaussian distributed.
As an example, we assume $n_k$ follows a skewed Beta distribution $B(\alpha, \beta)$, with $\beta=1.2\alpha$, and $\sigma^2_0=\frac{\alpha\beta}{(\alpha+\beta)^2(\alpha+\beta+1)}$.
From Fig. \ref{beta}, we observe that all the proposed detectors outperform the detector
optimized for the channel with Gaussian distributed $n_k$, and with a fixed offset of $b_k=\mu_b$.
The proposed DTD based on the the RNN detection achieves the performance of the optimum detector
with the full knowledge of the channel (with $n_k$ being Beta distributed), which is obtained
by simulations. This demonstrates that the proposed NN-based DTD works well for channels with the non-Gaussian distributed noise.

\section{Conclusions}
We have considered the memory physics induced unknown offset that severely degrades the error performance of many NVM channels, and proposed a novel NN-based dynamic threshold detection scheme. In particular, we have first proposed novel NN detectors, which can effectively tackle the unknown offset of the NVM channel. We found that the RNN detector outperforms the MLP detector, and approaches the performance of the optimum detector with the full knowledge of the channel. It also requires much smaller size of training data, and can learn the NVM channel uncertainty much faster than the MLP detector. However, compared with the conventional threshold detector, the NN detectors will result in a significant delay of the read latency and more power consumption. Therefore, we have further proposed a novel DTD, whose detection threshold can be derived based on the outputs of the proposed NN detectors. We proposed to only activate the NN-based detection when the ECC decoder fails, or periodically when the system is in the idle state. Thereafter, the threshold detector will still be adopted by using the adjusted detection threshold derived base on the outputs of the NN detector, until a further adjustment of the detection threshold is needed. Thus leading to a significant reduction of the read latency and power consumption. In the simulations, we have considered channels with both Gaussian distributed and non-Gaussian distributed noises. Simulation results demonstrated that the proposed DTD based on the RNN detection can achieve the error performance of the optimum detector, without the prior knowledge of the NVM channel. Thus demonstrating the great potential of the proposed NN-based DTD for NVMs.

\bibliographystyle{IEEEtran}
\bibliography{postdoc_refs}

\begin{thebibliography}{10}
\providecommand{\url}[1]{#1}
\csname url@samestyle\endcsname
\providecommand{\newblock}{\relax}
\providecommand{\bibinfo}[2]{#2}
\providecommand{\BIBentrySTDinterwordspacing}{\spaceskip=0pt\relax}
\providecommand{\BIBentryALTinterwordstretchfactor}{4}
\providecommand{\BIBentryALTinterwordspacing}{\spaceskip=\fontdimen2\font plus
\BIBentryALTinterwordstretchfactor\fontdimen3\font minus
  \fontdimen4\font\relax}
\providecommand{\BIBforeignlanguage}[2]{{%
\expandafter\ifx\csname l@#1\endcsname\relax
\typeout{** WARNING: IEEEtran.bst: No hyphenation pattern has been}%
\typeout{** loaded for the language `#1'. Using the pattern for}%
\typeout{** the default language instead.}%
\else
\language=\csname l@#1\endcsname
\fi
#2}}
\providecommand{\BIBdecl}{\relax}
\BIBdecl

\bibitem{yu2016emerging}
S.~Yu and P.-Y. Chen, ``Emerging memory technologies: recent trends and
  prospects,'' \emph{IEEE Solid State Circuits Mag.}, 2016.

\bibitem{chen2015portable}
B.~Chen, K.~Cai, G.~Han, S.~Lim, and M.~Tran, ``A portable dynamic switching
  model for perpendicular magnetic tunnel junctions considering both thermal
  and process variations,'' \emph{IEEE Trans. Magn.}, 2015.

\bibitem{cai2012flash}
Y.~Cai, G.~Yalcin, O.~Mutlu, E.~F. Haratsch, A.~Cristal, O.~S. Unsal, and
  K.~Mai, ``Flash correct-and-refresh: Retention-aware error management for
  increased flash memory lifetime,'' in \emph{In Proc. IEEE ICCD}, Sep. 2012.

\bibitem{papandreou2011drift}
N.~Papandreou, H.~Pozidis, T.~Mittelholzer, G.~Close, M.~Breitwisch, C.~Lam,
  and E.~Eleftheriou, ``Drift-tolerant multilevel phase-change memory,'' in
  \emph{In Proc. IEEE IMW}, May 2011.

\bibitem{wu2016temperature}
B.~Wu, Y.~Cheng, J.~Yang, A.~Todri-Sanial, and W.~Zhao, ``Temperature impact
  analysis and access reliability enhancement for \text{1T1MTJ}
  \text{STT-RAM},'' \emph{IEEE Trans. Rel.}, vol.~65, no.~4, pp. 1755--1768,
  2016.

\bibitem{lee2013estimation}
D.-h. Lee and W.~Sung, ``Estimation of nand flash memory threshold voltage
  distribution for optimum soft-decision error correction,'' \emph{IEEE Trans.
  Signal Process.}, vol.~61, no.~2, pp. 440--449, 2013.

\bibitem{schoeny2015analysis}
C.~Schoeny, F.~Sala, and L.~Dolecek, ``Analysis and coding schemes for the
  flash normal-laplace mixture channel,'' in \emph{Proc. IEEE ISIT}, Jun. 2015.

\bibitem{huang2014optimization}
K.~Huang, N.~Ning, and Y.~Lian, ``Optimization scheme to minimize reference
  resistance distribution of spin-transfer-torque mram,'' \emph{IEEE Trans.
  Very Large Scale Integr. (VLSI) Syst.}, 2014.

\bibitem{pelusi2015m}
D.~Pelusi, S.~Elmougy, L.~G. Tallini, and B.~Bose, ``$ m $-ary balanced codes
  with parallel decoding,'' \emph{IEEE Trans. Inf. Theory}, 2015.

\bibitem{immink2018composition}
K.~A.~S. Immink and K.~Cai, ``Composition check codes,'' \emph{IEEE Trans. Inf.
  Theory}, vol.~64, no.~1, pp. 249--256, 2018.

\bibitem{immink2014minimum}
K.~A.~S. Immink, J.~H. Weber \emph{et~al.}, ``Minimum pearson distance
  detection for multilevel channels with gain and/or offset mismatch.''
  \emph{IEEE Trans. Inf. Theory}, vol.~60, no.~10, pp. 5966--5974, 2014.

\bibitem{immink2018dynamic}
K.~A.~S. Immink, K.~Cai, and J.~H. Weber, ``Dynamic threshold detection based
  on pearson distance detection,'' \emph{IEEE Trans. Commun.}, 2018.

\bibitem{sutskever2014sequence}
I.~Sutskever, O.~Vinyals, and Q.~V. Le, ``Sequence to sequence learning with
  neural networks,'' in \emph{In Proc. NIPS}, Dec. 2014.

\bibitem{ye2018power}
H.~Ye, G.~Y. Li, and B.-H. Juang, ``Power of deep learning for channel
  estimation and signal detection in ofdm systems,'' \emph{IEEE Wireless
  Commun. Lett.}, vol.~7, no.~1, pp. 114--117, 2018.

\bibitem{gruber2017deep}
T.~Gruber, S.~Cammerer, J.~Hoydis, and S.~ten Brink, ``On deep learning-based
  channel decoding,'' in \emph{In Proc. IEEE CISS}, Mar. 2017.

\bibitem{kui2017cascaded}
K.~Cai and K.~S. Immink, ``Cascaded channel model, analysis, and hybrid
  decoding for spin-torque transfer magnetic random access memory
  (\text{STT-MRAM}),'' \emph{IEEE Trans. Magn.}, Nov. 2017.

\bibitem{mei2018magn}
Z.~Mei, K.~Cai, and B.~Dai, ``Polar codes for spin-torque transfer magnetic
  random access memory,'' \emph{IEEE Trans. Magn.}, 2018.

\bibitem{zhang2011stt}
Y.~Zhang, X.~Wang, and Y.~Chen, ``\text{STT-MRAM} cell design optimization for
  persistent and non-persistent error rate reduction: A statistical design
  view,'' in \emph{Proc. IEEE ICCAD}, Nov. 2011.

\bibitem{goodfellow2016deep}
I.~Goodfellow, Y.~Bengio, A.~Courville, and Y.~Bengio, \emph{Deep
  learning}.\hskip 1em plus 0.5em minus 0.4em\relax MIT press Cambridge, 2016,
  vol.~1.

\bibitem{zhong2018rate}
X.~Zhong, K.~Cai, P.~Chen, and Z.~Mei, ``Rate-compatible protograph \text{LDPC}
  codes for spin-torque transfer magnetic random access memory
  (\text{STT-MRAM}),'' in \emph{Proc. IEEE APMRC}, Nov. 2018.

\bibitem{mram}
\emph{Everspin Technologies MR4A16B Datasheet}.\hskip 1em plus 0.5em minus
  0.4em\relax Accessed: Mar. 27, 2018. [Online]. Available:
  https://www.everspin.com/supportdocs/MR4A16B\_ Datasheet.

\bibitem{keras}
F.~Chollet, ``keras,'' https://github.com/keras-team/keras, 2015.

\bibitem{tensor}
M.~Abadi \emph{et~al.}, ``Tensorflow: Large-scale machine learning on
  heterogeneous systems,'' 2015. [Online]. Available: http://tensorflow.org/.

\end{thebibliography}

\end{document}